\documentclass[10pt,conference]{IEEEtran}

\usepackage{tikz}
\usepackage{amsmath}
\usepackage{booktabs}
\usepackage{filecontents}
\usepackage{bbm}

\usepackage{xcolor}
\usepackage{enumitem}
\usepackage{graphicx}
\usepackage{subcaption}
\usepackage{hyperref}
\usepackage{xurl}
\usepackage{mathtools,amssymb,latexsym,amsfonts,stmaryrd}
\usepackage{amsmath}
\usepackage{listings}
\usepackage{multirow}
\usepackage{pifont}
\usepackage[many]{tcolorbox} %
\usepackage{listings}
\usepackage{xspace}
\usepackage{algorithm}
\usepackage[noend]{algpseudocode}
\usepackage{mathpartir}
\usepackage[inference]{semantic}
\usepackage{wrapfig}
\usepackage[capitalise]{cleveref}
\usepackage[numbers,sort&compress]{natbib}

\lstdefinestyle{base}{
  moredelim=**[is][\color{red}]{@}{@},
  escapeinside={<@}{@>}
}

\newcommand{\tool}{\textsc{HERTA}}

\newcommand{\F}{Fig.}

\renewcommand{\S}{Sec.}
\newcommand{\A}{Alg.}

\newcommand{\parh}[1]{\noindent\textbf{#1}}
\newcommand{\parhs}[1]{\noindent\textit{#1}}

\newcommand{\type}{\ensuremath{\tau}\xspace}
\newcommand{\btype}{\ensuremath{\varphi}\xspace}
\newcommand{\ptype}{\ensuremath{\omega}\xspace}
\newcommand{\reftype}[2]{\ensuremath{\{\nu: #1 \mid \nu: #2 \}}\xspace}
\newcommand{\idop}{\pi}
\newcommand{\binop}{\otimes}

\newcommand{\stmt}{\ensuremath{s}\xspace}

\newif\ifdebug
\debugfalse

\ifdebug
\newcommand{\modify}[1]{{\color{blue}#1}}

\else
\newcommand{\modify}[1]{#1}

\fi

\definecolor{codegray}{rgb}{0.95,0.95,0.95} %
\definecolor{codecomment}{rgb}{0.0,0.6,0.0} %
\definecolor{codekeyword}{rgb}{0.58,0,0.82} %
\definecolor{codestring}{rgb}{0.8,0.3,0.3}  %
\definecolor{codeblack}{rgb}{0.0,0.0,0.0}   %
\definecolor{ForestGreen}{rgb}{0.13,0.55,0.13} %

\usepackage[normalem]{ulem}

\begin{document}

\date{}

\title{Detecting and Understanding Vulnerabilities in Fully Homomorphic Encryption Frameworks}

\author{
\IEEEauthorblockN{
Yiteng Peng\IEEEauthorrefmark{1},
Dongwei Xiao\IEEEauthorrefmark{1},
Zhibo Liu\IEEEauthorrefmark{2},
Zhenlan JI\IEEEauthorrefmark{3},
Shuai Wang\IEEEauthorrefmark{1}
}
\IEEEauthorblockA{\IEEEauthorrefmark{1}The Hong Kong University of Science and Technology\\
\{ypengbp, dxiaoad, shuaiw\}@cse.ust.hk}
\IEEEauthorblockA{\IEEEauthorrefmark{2}State Key Laboratory of Novel Software Technology, Nanjing University\\
zhiboliu@nju.edu.cn}
\IEEEauthorblockA{\IEEEauthorrefmark{3}Nara Institute of Science and Technology\\
ji.zhenlan@naist.ac.jp}
}

\maketitle

\begin{abstract}

Fully homomorphic encryption (FHE) allows computations to be performed directly
on encrypted data without decryption, offering strong privacy guarantees for 
sensitive data analysis. This capability is important for
privacy-sensitive applications like secure cloud computing, finance, and
healthcare. The complexity of FHE schemes, however, has hindered their practical
adoption. To make FHE accessible to a broader range of developers, a new
generation of specialized frameworks has emerged to translate high-level FHE
programs into complex FHE operations, introducing a new programming paradigm.
However, the inherent complexity of FHE frameworks makes them prone to incorrect
implementation logic. Unlike mere crashes, logic bugs in these frameworks can
silently corrupt encrypted computation, potentially leading to severe financial
losses and security vulnerabilities in FHE-enhanced applications.

In this work, we introduce \tool, the first automated testing tool tailored for
FHE frameworks. \tool\ leverages metamorphic testing to uncover deep-seated
implementation bugs and vulnerabilities across the multi-layered FHE software
stack. To that end, we design a set of novel metamorphic relations (MRs) derived
specifically from FHE semantics. These MRs stress the most challenging aspects
of the pipeline, enabling automated correctness testing without the need for a
manual ground truth. 
Our evaluation of \tool\ on 3 leading industry frameworks discovered 21
previously unknown bugs, several of which have already been confirmed and fixed by
developers. Furthermore, our hazard analysis reveals the critical security
impact these bugs pose to the integrity and availability of FHE-based services.

\end{abstract}

\section{Introduction}
\label{sec:introduction}

Fully homomorphic encryption (FHE) has emerged as a transformative paradigm in
the privacy-preserving computation field~\cite{gentry2009fully,
cheon2017homomorphic, cheon2019full}. By enabling computations directly on
encrypted data without decryption, FHE addresses the long-standing dilemma
between data utility and privacy, facilitating critical applications in
privacy-preserving machine learning, secure genomic analysis, and confidential
cloud computing~\cite{wood2020homomorphic, wang2024secure}. With the maturation
of underlying FHE schemes, the demand for deploying FHE in real-world production
systems is growing rapidly.

Despite its promise, developing FHE-enhanced applications remains difficult for non-cryptographers. Developers must navigate complex
cryptographic parameters and manage ciphertext noise~\cite{bergerat2023parameter}. To bridge this gap, vendors like
Google and IBM have developed high-level FHE
frameworks~\cite{aharoni2023helayers, ali2025heir}. These frameworks take
high-level programs as input and automatically lower them into cryptographic
primitives or circuits, handling the heavy intermediate passes of parameter
selection, relinearization, and bootstrapping insertion.

While significantly lowering the adoption barrier, these frameworks
inevitably introduce an intricate multi-layered software stack. Since the
underlying cryptographic primitives lack native support for common non-linear
functions, such as \texttt{sine} and \texttt{maximum}, FHE frameworks perform
sophisticated adaptations to accommodate high-level inputs. Furthermore, to
mitigate the inherent performance overhead of homomorphic computation,
domain-specific optimizations, such as lookup table (LUT) fusion~\cite{Concrete}
and array layout permutation~\cite{aharoni2023helayers}, are crucial. The
complexity is further exacerbated by the need to interface with diverse
underlying cryptographic libraries~\cite{sealcrypto, openfhe} and varying
lower-level FHE schemes~\cite{cheon2017homomorphic, chillotti2020tfhe}. Given
the sensitive nature of input data and the critical domains served by
FHE~\cite{munjal2023systematic, he-fin5}, ensuring the correctness of these
frameworks is of paramount importance.

Despite these high stakes, automated testing for FHE frameworks remains largely unexplored and presents unique challenges. The FHE compilation and execution pipeline,
spanning frontend parsing, intermediate optimization, and backend cryptographic
execution, is highly complex. Moreover, bugs in FHE frameworks
often manifest as ``silent errors'' that subtly corrupt encrypted computations
rather than obvious failures like crashes or hangs. Such silent failures are
particularly insidious, as they can lead to significant 
security vulnerabilities in privacy-sensitive applications, as will be shown in
our hazard analysis in \S~\ref{subsec:security_implication}.

\begin{figure*}[t]
  \centering
  \includegraphics[width=0.9\linewidth]{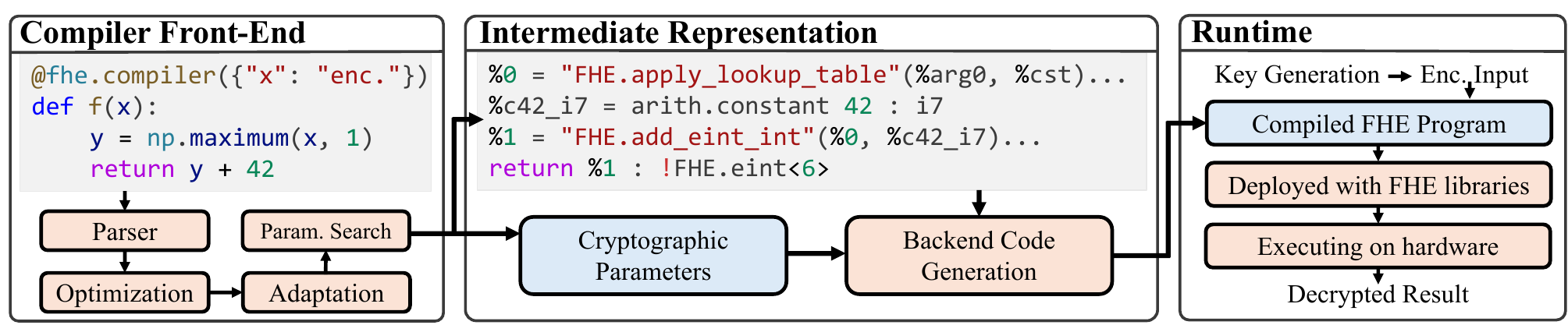}
  \caption{Overview of the pipeline in the FHE frameworks.}
  \label{fig:framework}
\end{figure*}

We introduce \tool, the first automated and systematic testing tool designed to
uncover bugs across the hierarchical stack of modern FHE
frameworks. \tool\ is designed on the basis of metamorphic testing (MT), a
powerful software testing scheme that checks program correctness through the
\textit{output consistency} under specific input transformations, known as
metamorphic relations (MRs)~\cite{chen2020metamorphic, segura2016survey}.
Specifically, we equip \tool\ with novel FHE-specific MRs that target
three critical stages of the compilation and execution stack: 1) frontend mutations:
targeting ciphertext types and dataflow structures to systematically transform
seed programs; 2) intermediate configurations: employing priority-driven
pairwise testing to achieve comprehensive coverage of the vast optimization
space despite high execution overhead; and 3) backend retargeting: implementing
equivalent transformations across disparate FHE schemes and hardware backends.
By applying these hierarchical MRs, \tool\ can detect both silent
logic errors and crashes that may expose new exploitation vectors,
which helps protect the integrity and availability of FHE-enhanced applications.

We implemented \tool\ and evaluated it on three leading FHE frameworks,
specifically Zama's Concrete~\cite{Concrete}, Google's HEIR~\cite{ali2025heir}, and
IBM's HELayers~\cite{aharoni2023helayers}. Our extensive testing campaign
successfully uncovered 21 bugs distributed across various layers of the
FHE frameworks, including 16 logic errors leading
to silent incorrect calculations and 5 crashes during
compilation and execution. %
We have reported these issues to the respective developers, with several of them already
confirmed. Furthermore, we conducted a pioneering hazard analysis on
the discovered bugs, demonstrating how these bugs can potentially compromise the
integrity and availability of real-world privacy-preserving applications.
In summary, we make the following contributions: 
\begin{itemize}[leftmargin=*,topsep=3pt,itemsep=2pt]
\item We advocate for a new initiative in security assurance for FHE frameworks,
identifying bugs that can silently corrupt computations or cause exploitable crashes in complex FHE frameworks.

\item We design \tool, an automated testing framework based on MT. It is
equipped with hierarchical, FHE-specific novel MRs and design
optimizations. It enables fine-grained and effective testing of the entire FHE
framework stack. 

\item We conduct a systematic evaluation on three major FHE frameworks
developed by industry, identifying 21 previously unknown bugs. Our analysis and exploitation further reveal that these bugs may constitute severe security hazards in real-world, privacy-preserving applications.
\end{itemize}

\section{Preliminary}
\label{sec:preliminary}

\subsection{Fully Homomorphic Encryption}

Homomorphic encryption (HE) enables computations on encrypted data without
accessing underlying plaintexts~\cite{rivest1978data}. For a function $f$
and sensitive input $x$, the client encrypts $\tilde{x} = \mathsf{Enc}(x)$, the
server evaluates $\tilde{y} = \tilde{f}(\tilde{x})$, and correctness guarantees
$\mathsf{Dec}(\tilde{y}) = f(x)$. Modern FHE schemes are based on hard 
problems like Learning With Errors (LWE)~\cite{regev2009lattices}. The
fundamental challenge is noise accumulation during homomorphic operations, which
grows with each operation and causes decryption failure when exceeding a
threshold. Gentry's \textit{bootstrapping}~\cite{gentry2009fully} refreshes
noisy ciphertexts by homomorphically evaluating the decryption circuit, though
at significant computational cost.

Modern FHE schemes target different data types and operations.
BFV~\cite{fan2012somewhat} and TFHE~\cite{chillotti2020tfhe} support exact
arithmetic on modular integers and bits, respectively, while
CKKS~\cite{cheon2017homomorphic} enables approximate real-number computations
for machine learning. TFHE's programmable bootstrapping (PBS) evaluates
non-linear functions via lookup tables during noise
refresh~\cite{lou2020glyph, guimaraes2021revisiting}. BFV and CKKS use SIMD
packing to encode plaintext vectors into ciphertext slots, requiring expensive
rotation operations for cross-slot dependencies~\cite{ran2023spencnn}.

\subsection{FHE Frameworks}

While promising for privacy, implementing applications directly on raw
cryptographic primitives is difficult as it requires manual handling of complex
ring arithmetic and polynomial operations. Libraries such as SEAL~\cite{sealcrypto}, OpenFHE~\cite{openfhe}, and TFHE-rs~\cite{TFHE-rs} alleviate this problem to some extent by implementing common FHE primitives like addition, multiplication, and rotation for various FHE schemes, such as BFV, CKKS, and TFHE. However, developers are still required to possess deep cryptographic expertise to select secure encryption parameters, manually manage noise growth, and adapt non-linear functions using a limited set of supported operations. Improper configuration can lead to decryption failures or performance degradation of several orders of magnitude.

To democratize FHE, industry leaders developed high-level frameworks like
Google's HEIR~\cite{ali2025heir}, Zama's Concrete~\cite{Concrete}, and
IBM's HELayers~\cite{aharoni2023helayers}, abstracting underlying FHE
libraries.
\cref{fig:framework} demonstrates a high-level workflow of FHE frameworks.\footnote{We will use the term ``compiled program'' to refer to the output of FHE
frameworks, which will be API calls to underlying FHE libraries in our context. The term ``runtime execution'' refers to the execution of API calls on FHE libraries.}
These frameworks accept high-level
programs, such as Python, MLIR, or domain-specific APIs, as input and
automatically lower them into optimized invocations of low-level FHE
primitives in the FHE libraries, such as OpenFHE or SEAL. By abstracting away the underlying cryptographic complexities, they allow developers to focus on core business logic. 
Since FHE operations are $10^3\times$ to $10^6\times$ slower than plaintext
execution \cite{brynds2025cryptoracle,peng2025testing}, these frameworks apply
crucial optimizations like array layouts, LUT fusion, and bootstrapping
scheduling to enhance efficiency.

FHE frameworks perform automatic analysis to synthesize cryptographic parameters
satisfying security requirements while accommodating noise constraints. Since
user programs exhibit diverse computational depth and characteristics, static
policies are insufficient. Modern frameworks employ sophisticated techniques to
precisely estimate noise accumulation and optimize bootstrapping scheduling,
balancing between conservative strategies that guarantee correctness but incur
prohibitive performance penalties and aggressive optimizations that maximize efficiency.

Beyond parameter selection, FHE frameworks bridge the semantic gap between
high-level program semantics and restricted FHE arithmetic. This includes
mapping native data types to cryptographic representations (for example,
Concrete-ML~\cite{ConcreteML} quantizes floating-point values for
TFHE~\cite{chillotti2020tfhe}) and approximating unsupported operators like
non-linear functions via polynomial
approximation~\cite{chen2018logistic} or LUTs~\cite{lou2020glyph}. These
adaptations create a vast design space for trading off accuracy and efficiency.
Frameworks also implement multi-layered optimizations, from generic passes like
dead code elimination to domain-specific transformations such as tensor layout
optimization~\cite{aharoni2023helayers} and LUT fusion~\cite{Concrete}.

Note that modern FHE frameworks adopt modular designs supporting backend
retargetability, allowing programs to target different hardware like CPUs and GPUs, various cryptographic libraries like SEAL~\cite{sealcrypto} and OpenFHE~\cite{openfhe}, or FHE schemes like BFV~\cite{fan2012somewhat} and BGV~\cite{brakerski2014leveled}. While this flexibility enhances portability,
it creates a deep and heterogeneous software stack where each layer, from
semantic bridging and parameter selection to optimization and code
generation, introduces distinct failure possibilities.

\subsection{Metamorphic Testing}
Traditional testing relies on test oracles (ground truth) to verify output correctness. However, in complex domains like compilers and cryptography, determining correct outputs is often
intractable, a challenge known as the oracle problem~\cite{barr2014oracle}.
Metamorphic testing (MT)~\cite{chen2020metamorphic} addresses this by verifying
invariant properties termed \textit{metamorphic relations} (MRs).

Formally, given a source test case $I_{s}$ and transformation $\mathcal{T}$,
MT generates a follow-up test case $I_{t} = \mathcal{T}(I_{s})$. The system
executes both to obtain outputs $O_{s}$ and $O_{t}$, and an MR
prescribes the expected relation $\mathcal{R}$ between these outputs. Any
violation indicates a potential defect, bypassing the need for reference
implementations.

\noindent\textbf{Example.} Consider testing a sine function implementation. We
can use the mathematical property $\sin(x) = \sin(\pi-x)$ as an MR. Given source
input $I_{s}=x$, we apply transformation $\mathcal{T}(x) = \pi-x$ to generate
follow-up input $I_{t}=\pi-x$. After computing $O_{s} = \sin(I_{s})$ and
$O_{t} = \sin(I_{t})$, the MR requires $O_{s} = O_{t}$. If this
equality fails, the sine implementation contains a bug.

\begin{figure*}[t]
  \centering
  \lstset{
    language=Python,
    basicstyle=\ttfamily\scriptsize,
    frame=single,
    numbers=none,
    xleftmargin=0pt,
    framesep=2pt,
    aboveskip=1pt,
    belowskip=1pt,
    breaklines=true,
    columns=fullflexible
  }

  \begin{subfigure}[t]{0.32\textwidth}
\begin{lstlisting}
scalar = fhe.constant(4)
arr = fhe.array([0, 1, x, 2, 3])
z = scalar * arr
return z[2] + x
\end{lstlisting}
    \caption{Logic error in high-level program lowering. The Concrete framework mishandles the dataflow when mixing encrypted inputs with array packing, leading to a silent bug.}
    \label{fig:mot_concrete_array}
  \end{subfigure}
  \hfill
  \begin{subfigure}[t]{0.32\textwidth}
\begin{lstlisting}
t = np.abs(-2) | np.abs(15)
v = np.maximum(x, -13 + y)
w = (v >= t)
return t + w + x
\end{lstlisting}
    \caption{Assertion failure triggered by optimization choice. Enabling
        the specific \texttt{THREE\_TLU\_CASTED} strategy causes an internal
        compiler error on valid code.}
    \label{fig:mot_concrete_config}
  \end{subfigure}
  \hfill
  \begin{subfigure}[t]{0.32\textwidth}
\begin{lstlisting}
# backend = SealCkksContext ()
v = 12 + (5 - x)
w = -14 - (-7 - v)
return w + (-7 - v)
\end{lstlisting}
    \caption{Semantic divergence across backends. The identical HELayers code yields correct results on OpenFHE but fails silently on the SEAL backend.}
    \label{fig:mot_helayers}
  \end{subfigure}

    \caption{Simplified motivating examples discovered by \tool. These cases demonstrate that vulnerabilities permeate the entire multi-layered FHE framework, from high-level code lowering (a) and optimization configuration (b) to backend switching (c).}
  \label{fig:motivation_examples}
\end{figure*}

\section{Motivation}
\label{sec:motivation}

\subsection{Vulnerability and Significance}
\label{subsec:motivation_significance}

FHE has transitioned from a theoretical construct to a cornerstone of
privacy-preserving computation, driven by major industrial frameworks such as
Zama's Concrete, Google's HEIR, and IBM's HELayers. With increasing adoption
in high-stakes domains like finance and
healthcare~\cite{munjal2023systematic,he-fin5}, the correctness of their implementation becomes
paramount. Unlike traditional crashes, vulnerabilities in FHE frameworks often
manifest as ``silent errors'', which are logic bugs that corrupt encrypted results without
outward failure. 
The real-world impact of such errors can be catastrophic: in finance, a silent
miscalculation during encrypted auditing could lead to manipulated credit assessments while the source code appears %
semantically correct; in healthcare, it could lead to incorrect medical
diagnoses based on corrupted genomic analysis, %
directly endangering patient safety. Given these significant economic and safety
risks, systematically detecting these logic flaws is necessary to safeguard the integrity of privacy-sensitive applications.

\subsection{Research Challenges}

Testing FHE frameworks poses unique challenges compared to traditional
compiler testing, stemming from the novel cryptographic characteristics discussed in
\S~\ref{sec:preliminary}. To make this powerful paradigm accessible to
developers lacking deep cryptographic expertise, modern FHE frameworks abstract
away the underlying mathematical complexity. They attempt to encapsulate
intricate computations within the rigid constraints of cryptographic primitives.
This abstraction requires a complex compilation process, which must
not only translate familiar high-level constructs into low-level arithmetic
circuits but also meticulously manage properties unique to FHE, most notably
ciphertext noise growth.

These cryptographic constraints introduce tight coupling and potential failure
modes at every level of the framework's hierarchical stack, from the high-level
language frontend to the backend scheme selection. Traditional compiler testing
or analysis methods, however, are ill-equipped to systematically explore this
intricate space, as they lack the domain awareness necessary to reason about
such FHE-specific, cross-layer concepts. Below, we provide three concrete
examples to illustrate fundamental testing challenges: logic errors in high-level
lowering, complex configuration interactions, and cross-backend semantic
inconsistencies.

\subsection{Challenge I: High-Level Program Lowering}
\label{subsec:motivation-challenge1}

The first challenge in testing FHE frameworks stems from two fundamental
properties of the underlying cryptography. As mentioned in
\S~\ref{sec:preliminary}, a unique challenge of FHE is the management of
\textit{ciphertext noise}, where excessive noise accumulation can render
ciphertexts undecryptable. Therefore, an essential responsibility of an FHE
framework is to manage the \textit{noise budget} by synthesizing appropriate
encryption parameters and strategically inserting bootstrapping operations to
reset noise levels. Furthermore, the framework must bridge the gap between
high-level data structures and low-level polynomial ring representations via
correct and efficient encoding. This involves non-trivial transformations, such
as representing floating-point numbers on integer rings through scaling, or packing multiple
data elements into single ciphertexts to maximize throughput. The interplay
between data representation and noise management further creates complex
dependencies that the framework must navigate correctly.

This complexity often manifests as silent bugs where an incorrect implementation
leads to precision loss or corrupted results without obvious manifestations.
Generating test cases that can systematically stress these interconnected
properties is a crucial yet challenging task for uncovering deep,
lowering-related vulnerabilities. Consider the program from the Concrete
framework~\cite{Concrete} in \cref{fig:mot_concrete_array}: the program
attempts to construct an array by mixing an encrypted scalar input $x$ with
plaintext constants in line 2. This operation forces the framework to handle
data layout transformations, while managing the broadcasting of operations in
line 3. Although the framework handles the array creation and element extraction correctly, it fails to maintain semantic consistency when the
extracted value is added back to the original input $x$. This bug exemplifies
the difficulty in managing hybrid dataflows and packing strategies. Without
targeted test cases that specifically exercise these
aspects, such logic errors about computational depth and data encoding would
remain hidden.

\subsection{Challenge II: Configuration Complexity}
\label{subsec:motivation-challenge2}

Unlike traditional compilation, where most operations can be deterministically
mapped to assembly instructions, high-level FHE operations, such as comparison,
maximum, ReLU, lack direct primitive equivalents in the underlying cryptographic
schemes. Instead, FHE frameworks must synthesize these operations into sequences
of supported primitives. This synthesis is not unique; a single high-level
operator can often be lowered via multiple distinct strategies. For instance, a
ReLU function can be approximated using a polynomial~\cite{lee2022low}, or evaluated
precisely via LUTs~\cite{nandakumar2019towards}. Each lowering strategy offers distinct
trade-offs in accuracy, performance, and noise consumption, and frameworks often
provide users with configuration options to select among these strategies.

This configurability creates a combinatorial explosion of
compilation paths where specific settings can interact unexpectedly with program
constructs. Consequently, effective testing requires more than valid input
generation; it necessitates a systematic exploration of this vast configuration
space. The code in \F~\ref{fig:mot_concrete_config} demonstrates a subtle bug
arising from such configuration interactions. It executes successfully by
default but crashes with an internal assertion failure when the
\texttt{min\_max\_strategy\_preference} is set to \texttt{THREE\_TLU\_CASTED}, i.e.,
instructing the framework to synthesize the \texttt{min} and \texttt{max}
operations using three LUTs. This demonstrates how valid programs can fail due
to specific configuration-dataflow interactions. Since exhaustively exploring
this combinatorial space is prohibitively expensive, we need a prioritized
testing strategy to navigate its complexities.

\subsection{Challenge III: Inconsistent Backend/Scheme}
\label{subsec:motivation-challenge3}

Modern FHE frameworks are designed to be modular, often supporting multiple
cryptographic schemes, such as BGV, CKKS, and TFHE, and backend libraries, including
OpenFHE~\cite{openfhe}, SEAL~\cite{sealcrypto}, HElib~\cite{halevi2020design}.
As discussed in \S~\ref{sec:preliminary}, FHE schemes differ fundamentally in
their capabilities; some, like BGV, perform exact integer arithmetic, while
others, like CKKS, perform approximate arithmetic on real numbers. Furthermore,
different backend libraries encapsulate FHE primitives differently,
exposing unique parameters and exhibiting distinct performance and noise
management characteristics even when implementing the same scheme.

While this modularity offers flexibility, it creates unique challenges in
finding subtle bugs that lead to cross-stack inconsistencies. Na\"ively
cross-checking outputs across different backends or schemes is insufficient, as
these components have inherent semantic differences. For instance, checking
consistency between an exact integer scheme and an approximate arithmetic scheme
requires a nuanced oracle that tolerates mathematical divergence. Such semantic
gaps complicate testing oracle design. Moreover, translating a single high-level
program into correctly configured primitives for different backends is a complex
and error-prone process.

\cref{fig:mot_helayers} illustrates a semantic divergence we discovered in
HELayers. A valid CKKS program executes correctly with OpenFHE but silently
computes incorrect results on the SEAL backend. This discrepancy highlights the
difficulty of ensuring semantic preservation across heterogeneous cryptographic
foundations. Uncovering these bugs necessitates systematically exploring the
combinatorial space of backend and scheme configurations to detect such subtle
but critical inconsistencies. The cross-product
of multiple schemes, diverse backend libraries, and varying hardware targets
creates a vast testing surface where semantic preservation is not guaranteed and
traditional strict-equality oracles are hardly applicable.

\section{Methodology}
\label{sec:design}

\noindent\textbf{Study Scope.}
To systematically address challenges in \S~\ref{sec:motivation}, we
introduce \tool, a testing tool tailored for FHE frameworks.
In general, \tool\ exposes two categories of implementation defects:

\begin{itemize}[leftmargin=1em]

    \item \textit{Logic Errors.} These bugs occur when the program executes to
    completion, but the decrypted result is mathematically incorrect. Such
    errors often stem from the intricate compilation and runtime pipeline, where the
    complexities of data representation, noise management, and scheme-specific
    optimizations can lead to silent miscompilations. Given that FHE frameworks
    are increasingly deployed in privacy-sensitive domains
    (\S~\ref{subsec:motivation_significance}), these silent errors are critical,
    as they undermine the computation and lead to flawed usage.

    \item \textit{Crashes.} This category encompasses explicit failures where
    the FHE framework or the compiled program terminates abnormally. These
    defects typically manifest as internal assertion failures or memory
    corruption events that occur either during the complex compilation phases or
    at runtime.

\end{itemize}

\noindent\textbf{Main Users.} \tool\ is designed as a defensive tool for FHE framework developers and vendors, not intended for malicious exploitation. Our work assists them in testing their frameworks before release and during maintenance. As shown in \S~\ref{subsec:eval_overall}, major framework developers have confirmed our approach's utility by responding to our bug reports. \tool\ uncovers subtle logic errors, particularly silent miscompilations, that are difficult to detect with existing techniques. These findings are highly critical, as they often indicate potential
vulnerabilities that could be exploited in deployed privacy-preserving
applications~\cite{yadavalli2025homomorphic, smart2023practical}.

\noindent\textbf{Formulation.} We formalize the behavior of an FHE framework as
a function $\mathcal{F}: \mathcal{P} \times \mathcal{C} %
\rightarrow \mathcal{R}$, where $\mathcal{P}$ denotes the space of valid input
programs, $\mathcal{C}$ represents the configuration space, and $\mathcal{R}$ is
the space of decrypted results. For a given program $P \in \mathcal{P}$ and
configuration $\Gamma \in \mathcal{C}$, the framework execution $\mathcal{F}(P,
\Gamma)$ produces a result $R \in \mathcal{R}$. We model a framework configuration as $\Gamma = (\sigma, \beta,
\textit{C}_{crypto}, \textit{C}_{ada}, \textit{C}_{opt})$, which comprises the following components: the FHE
scheme $\sigma$ (such as BGV, CKKS, or TFHE), the backend implementation $\beta$
(for example, SEAL, OpenFHE, or hardware targets like GPU), the cryptographic
parameters $\textit{C}_{crypto}$ (including polynomial modulus degree), the adaptation strategies $\textit{C}_{ada}$ for non-native operations
(such as the choice of approximation method for non-linear functions), and the
optimization settings $\textit{C}_{opt}$ (for instance, LUT fusion). This
formalization enables defining MRs as transformations on either
the program space $\mathcal{P}$ or the configuration space $\mathcal{C}$,
establishing rigorous equivalence conditions for detecting framework defects.

\begin{figure}
  \centering
  \includegraphics[width=1\linewidth]{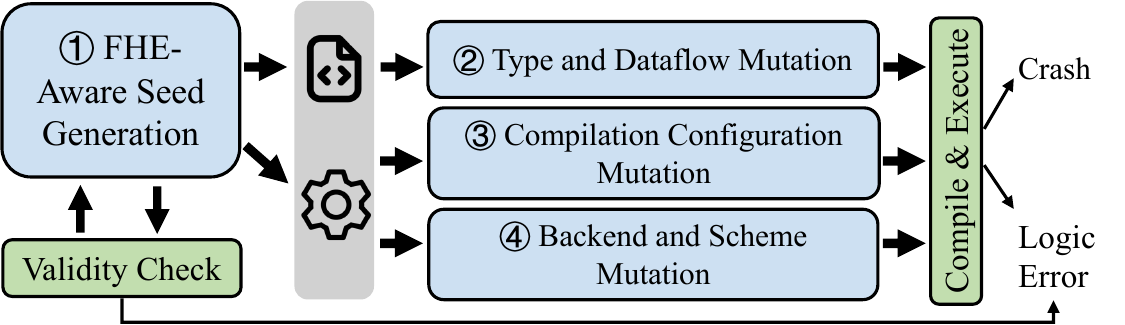}
  \caption{Overview of \tool's testing pipeline}
  \label{fig:pipeline}
\end{figure}

\noindent\textbf{Pipeline Overview.} An overview of \tool's pipeline is
illustrated in \cref{fig:pipeline}. The pipeline consists of four main
components: an FHE-aware seed generator and three mutation stages that mutate the
seed program with three MRs, respectively.

\parhs{\ding{192} FHE-aware Seed Generation.} This stage generates a seed
program, $P$, with FHE-specific constructs, such as complex data types and
operations requiring significant adaptation. This seed program will be used as
the basis for subsequent metamorphic mutations to stress-test the FHE framework.

\parhs{\ding{193} MR1: Type and Dataflow Mutation.} This MR transforms $P$ into a
semantically equivalent $P'$ with FHE-specific mutations, checking if
$\mathcal{F}(P, \Gamma) = \mathcal{F}(P', \Gamma)$. It targets
FHE-specific data types and expression structures to test the framework's
management of encrypted data and computational flow.

\parhs{\ding{194} MR2: Compilation Configuration Mutation.} This component
systematically mutates compilation configurations, such as cryptographic
parameters $\textit{C}_{crypto}$, the adaptation strategies $\textit{C}_{ada}$
and optimization settings $\textit{C}_{opt}$. It uses prioritized combinatorial
testing to effectively explore the vast configuration space and test the
framework's robustness across diverse settings.

\parhs{\ding{195} MR3: Backend and Scheme Mutation.} This MR tests for
cross-stack inconsistencies by mutating the FHE scheme $\sigma$ and the backend
$\beta$. By validating the equivalence of results across different cryptographic
foundations, it ensures correctness across the heterogeneous FHE ecosystem,
including software implementations and hardware accelerators.

For each generated test case, which consists of a seed program $ P $ and a related
configuration $ \Gamma $, \tool\ executes it and its mutated variants through
the FHE framework under test. Any explicit failure, such as a segmentation fault
or an assertion failure during either process, is immediately reported as a
crash. If all executions are successful, their decrypted results are compared. A
discrepancy between their outputs reveals a logic error.

\subsection{FHE-aware Seed Generation}
\label{subsec:seed_generation}

\parh{Design Goal.} The primary objective of seed generation is to produce a
corpus of initial FHE programs $P$ that are not only syntactically and
semantically valid but also specifically crafted to stress the FHE framework
stack. In contrast to traditional compiler testing, our process prioritizes the
generation of diverse, FHE-specific constructs. This includes deliberately
generating programs with deep multiplication paths to challenge noise
management, as multiplication operations significantly contribute to noise
growth in FHE and thus holistically test the framework's noise analysis and cryptographic parameter selection logic. Additionally, the
programs incorporate a rich variety of operations that lack direct mappings to
FHE primitives, such as comparisons and non-linear functions, which require
complex adaptation strategies.

\parh{Grammar of FHE Programs.} To systematically guide this process, we
define a context-free grammar, $\Lambda_{\tool}$, which abstracts the
syntax of high-level FHE programming languages. $\Lambda_{\tool}$ is not designed
to cover every syntactic detail of a specific FHE framework. Instead, it is
intentionally simple yet expressive enough to capture the essential FHE programming features, ensuring that generated programs are adaptable
across different frameworks. \F~\ref{fig:ir} shows the selected syntax of $\Lambda_{\tool}$. The grammar explicitly distinguishes two concepts
critical to FHE: the privacy type $\ptype$, which separates plaintext from
encrypted values, and the indirect operators $\idop$, which represent
high-level operations lacking direct support from FHE
primitives and thus requiring adaptation.

\begin{figure}[!htbp]
  \begingroup
  \small %
  \[
  \begin{array}{lccl}
    \emph{Program} \quad & P &::=& s; P \mid s\\
    \emph{Statement} \quad & \stmt &::=& assign \mid \emph{other structure} \\
                                                & & &\emph{supported by FHE frameworks} \\
    \emph{Assignment} \quad & assign &::=& v := e \\
    \emph{Variable} \quad & v & ::= & \emph{valid variable name} \\
    \emph{Constant} \quad & c &::=& \emph{valid values of type \btype} \\

    \emph{BasicOp} \quad & \binop &::=& + \mid - \mid \times \\
    \emph{IndirectOp} \quad & \idop &::=& \emph{operators not directly supported} \\
    \emph{BasicType} \quad & \btype &::=& \texttt{int} \mid \texttt{float} \\ %
    \emph{PrivacyType} \quad & \ptype &::=& \emph{plaintext} \mid \emph{encrypted} \\
    \emph{Type} \quad & \type &::=& \reftype{\btype}{\ptype} \\
    \emph{Expression} \quad & e &::=& c \mid v \mid e_1\xspace \binop \xspace e_2 \mid \idop(e_1, \dots) \\
  \end{array}
    \]
  \endgroup
  \caption{Selected abstract grammar for FHE programs.}
  \label{fig:ir}
\end{figure}

\parh{Program Generation Algorithm.} The generation of a program begins with the
start symbol $P$ and recursively expands non-terminal symbols by replacing them
with symbols from the right-hand side of applicable production rules, continuing
until only terminal symbols remain. For non-terminals with multiple production
options, the choice is made based on dynamically adjusted probability weights
that prioritize FHE-specific constructs. Specifically, the generator maintains a
context tracking the current circuit depth and the distribution of
operator types. As the program grows, probability weights are dynamically
adjusted. Initially, the generator biases towards arithmetic operations (such as
multiplication) to rapidly accumulate multiplicative depth; as the
multiplicative depth reaches a configurable threshold, the bias shifts towards
framework-specific indirect operations (such as comparisons) to increase
functional diversity without exceeding cryptographic constraints. This strategy
stresses the noise management of FHE frameworks while enriching seed diversity
for subsequent metamorphic mutations. To ensure seed validity, we adopt
lightweight self-checking during framework execution: seeds that cause
compilation or runtime failures (such as exceeding the noise budget) are
discarded and regenerated, whereas unexpected crashes are reported as potential
bugs. When the framework provides a plaintext reference, the seed's decrypted
result will also be checked.

\A~\ref{al:expr_generation} outlines the core logic of our expression generation
process. After \textsc{AdjustOpWeights} in line 2 dynamically calibrates
operation probabilities as discussed above, the \textsc{GenExpr} function
invokes \textsc{SelectOp} to determine the next operation type. This selection
is driven by dynamically adjusted weights in line 3 that prioritize arithmetic
operations early to accumulate depth, while shifting towards indirect operations
$\idop$ later to maximize feature coverage. Subsequently, operands are chosen
via a dual-mode heuristic. The generator stochastically toggles between
selecting high-depth variables in line 7 to stress noise management and
infrequently used variables in line 9 to foster diverse combinations of
multiplication depths. Finally, the symbol table $\Sigma$ and annotation set
$\mathcal{A}$ are updated to reflect the new state, enabling context-sensitive
decisions in subsequent expressions. Upon completing the recursive construction
of program $P$, \tool\ appends a default configuration $\Gamma$, which results
in a complete executable seed $(P, \Gamma)$ and serves as the foundation for the
MRs below. 

\begin{algorithm}
\caption{Expression Generation Strategy}
\label{al:expr_generation}
\footnotesize
\begin{algorithmic}[1]
\Function{GenExpr}{Symbol Table $\Sigma$, Annotation Set $\mathcal{A}$, Max Depth $D_{max}$}
    \State $W$ $\gets$ \Call{AdjustOpWeights}{$\mathcal{A}$, $D_{max}$}
    \State $op$ $\gets$ \Call{SelectOp}{$W$}
    \State $\mathcal{A}$ $\gets$ \Call{UpdateAnnotation}{$\mathcal{A}$, $op$} \Comment{Record used operations}
    \For{$i$ in $1 \dots \Call{NumOpExpr}{op}$}
        \If{$\Call{ChooseMaxDepthVar}{ }$}
            \State $opExpr_i \leftarrow \Call{GetMaxDepthVar}{\Sigma}$ 
        \Else
            \State $opExpr_i \leftarrow \Call{GetLeastUsedVar}{\Sigma}$
        \EndIf
        \State $\Sigma[opExpr_i].\text{usage} \leftarrow \Sigma[opExpr_i].\text{usage} + 1$
    \EndFor
    \State $expr$ $\gets$ $op$($opExpr_1, \dots$)
    \State \Return $expr$, $\mathcal{A}$
\EndFunction
\end{algorithmic}
\end{algorithm}

\subsection{MR1: Type and Dataflow Mutation}
\label{subsec:mr1}

\parh{Design Goal.}
This MR addresses Challenge I (\S~\ref{subsec:motivation-challenge1}) by stressing
the FHE framework's implicit data layout and noise management. It systematically
mutates the seed program $P$ into equivalent variants $P'$ with data-centric
transformations that directly impact how data is represented and how computations
are orchestrated in the encrypted domain. It then validates whether
$\mathcal{F}(P, \Gamma) = \mathcal{F}(P', \Gamma)$. Specifically, this MR
mutates programs with two complementary focuses. It modifies data types and
layouts, such as variable bit-widths and tensor shapes, to stress the framework's
data packing and representation logic. It also reshapes the dataflow structure of
the computation graph to challenge the framework's noise management capability.

\parh{Mutation Operators.} To systematically explore the program space sensitive
to FHE properties, we introduce a set of specialized mutation operators divided
into two categories: type mutations and dataflow mutations.\footnote{Here
``type'' refers to data types and layouts in the FHE context.}

\parhs{Type mutations.} This category stresses the framework's ciphertext data
representation and packing logic. Key operators include: \ding{192} \textit{Data
type alteration}, which promotes variable types and bit-widths (such as
\texttt{Secret[i16]} $\rightarrow$ \texttt{Secret[i32]}) to test the framework's
parameter selection logic under varying data range constraints; \ding{193}
\textit{Data layout mutation}, which systematically transforms scalar operations
into vectorized tensor computations. 
For example, this operator can employ a
``wrap-compute-unwrap'' pattern, transforming $x \times y$ into
$\bigl($$\texttt{[1,}$$x$$\texttt{,2]}$$ \times y\bigr)$$\texttt{[1]}$. This embeds scalars into
arrays before computation and then extracts the corresponding result, forcing the FHE framework
to exercise complex
ciphertext packing and index resolution paths; and \ding{194}
\textit{Plaintext-ciphertext swapping}, which toggles the encryption status of
operands between plaintext and ciphertext representations to verify the
correctness of arithmetic involving mixed privacy levels. Although simple to
implement, these transformations can trigger diverse optimization paths by
forcing the framework to re-synthesize cryptographic parameters and adapt data
packing strategies.

\parhs{Dataflow mutations.} This category targets the computational graph
structure to challenge the framework's noise management and optimization
capabilities. We design the following mutators: \ding{195} \textit{Expression
reshaping}, which transforms algebraic expressions, such as rewriting $v \times
v$ as $v^2$, or $v + v$ as $2 \times v$. This transformation potentially alters
multiplication depth and exercises different lowering paths; \ding{196}
\textit{Redundant operation insertion}, which injects redundant
arithmetic operations in the ciphertext domain, such as transforming $x + y$ into
$(x + 0_{\mathsf{enc}}) + (y \times 1_{\mathsf{enc}})$. This preserves the
mathematical outcome while forcing the framework to manage additional
noise accumulation, which stresses the framework's noise estimation accuracy and management
capability; and \ding{197} \textit{Equivalent non-linear insertion},
which replaces a variable with an equivalent non-linear function, such as $v
\rightarrow \max(v, \texttt{MIN\_VAL})$. As the transformed expression cannot be
directly mapped to FHE primitives, the framework must invoke complex adaptation strategies, thereby testing the adaptation layer.

\subsection{MR2: Compilation Configuration Mutation}
\label{subsec:mr2}

\parh{Design Goal.} This MR addresses Challenge II
(\S~\ref{subsec:motivation-challenge2}) by systematically mutating framework
configurations $ \Gamma $. Unlike MR1, which alters the program $P$, this relation focuses
on the compilation configurations of FHE frameworks. Specifically, it mutates the
adaptation strategies $\textit{C}_{ada}$ that control the lowering of high-level
non-linear operations, the optimization settings $\textit{C}_{opt}$ that govern
passes like subexpression elimination, and the cryptographic parameters
$\textit{C}_{crypto}$ such as polynomial modulus degree when applicable.
Note that this MR does not mutate the scheme $\sigma$ or backend $\beta$,
which are handled by MR3. Regardless of the specific configuration $\Gamma'$
applied, a correct FHE framework should yield consistent results, i.e.,
$\mathcal{F}(P, \Gamma) = \mathcal{F}(P, \Gamma')$.

\parh{Technical Challenge.} As highlighted in
\S~\ref{subsec:motivation-challenge2}, the configuration space of FHE frameworks is
combinatorially large, with dozens of configurable parameters that interact in
subtle and non-obvious ways. For instance, a high-level comparison operation can
be lowered using multiple adaptation strategies, each presenting different
trade-offs in performance, precision, and noise management. When compounded with
global optimization flags or adjustable cryptographic parameters when applicable,
the number of possible configuration combinations can easily scale into 
thousands or tens of thousands. Exhaustively executing all configurations is
computationally infeasible due to the high cost of FHE execution (often several
orders of magnitude greater than plaintext). Na\"ive random sampling also proves
inefficient, often wasting resources on configurations that do not impact the
given program's behavior. This necessitates a targeted strategy to efficiently
navigate promising regions of the configuration space.

\parh{Prioritized Combinatorial Testing.} To address the challenge of
combinatorial explosion, \tool\ employs a two-pronged strategy that integrates
systematic coverage of combinatorial testing with a diversity-guided
prioritization heuristic. This approach efficiently navigates the unified
parameter space, spanning $C_{ada}$, $C_{opt}$, and $C_{crypto}$, by ensuring
diversity of the selected configurations.

The foundation of our strategy is pairwise (2-way) combinatorial testing. As a
widely adopted method~\cite{nie2011survey}, this technique is proven to
effectively detect interaction-triggered faults while significantly reducing the
configuration search space by generating a candidate suite that covers every
possible pair of parameter values. We choose 2-way coverage as it provides a
practical balance: empirical studies show that most configuration bugs involve
relatively few parameter interactions~\cite{kuhn2004software, kuhn2009combinatorial}, while higher-way coverage incurs
prohibitive costs. However, given the high computational cost of FHE execution,
running the full pairwise suite remains impractical. To address this, \tool\
further refines the test suite through a usage-aware pruning mechanism followed
by diversity-guided selection.

To prune parameters that may not influence the program's behavior, \tool\
excludes parameters that do not correspond to any operators present in the seed
program $P$. For instance, if a seed program contains only linear arithmetic,
there is no need to test adaptation strategies for non-linear operations like
LUT configurations. This prevents wasting resources on logic paths that are
never exercised. Another prioritization heuristic is based on maximizing
configuration diversity. When selecting the next configuration from the candidate
suite, \tool\ selects the one that maximizes distance from previously executed
configurations. By prioritizing configurations that differ significantly from
prior tests, \tool\ increases the likelihood of uncovering corner cases where
multiple configurations interact in unexpected ways.

\begin{algorithm}[t]
\caption{Prioritized Configuration Testing}
\label{alg:prioritized_config_testing}
\footnotesize
\begin{algorithmic}[1]
\Function{PriConfigMT}{Program $P$, Initial Configuration $\Gamma$, Max Tests $N_{max}$}
\State $R \gets \Call{Execute}{P, \Gamma}$ %
\State $Ops$ $\gets$ \Call{RetrieveOperators}{$P$}
\State $Params$ $\gets$ \Call{GetRelevantParameters}{$Ops$} %
\State $ParamList$ $\gets$ \Call{GenPrioritizeSuite}{$Params$, $\Gamma$, $N_{max}$} %
\For{each $\Gamma'$ in $ParamList$}
\State $R' \gets \Call{Execute}{P, \Gamma'}$
\If{Execution Failed}
\State \Call{ReportPotentialCrash}{ }
\EndIf
\If{$R' \neq R$}
\State \Call{ReportInconsistency}{ }
\EndIf
\EndFor
\State \Return
\EndFunction
\end{algorithmic}
\end{algorithm}

\A~\ref{alg:prioritized_config_testing} outlines this prioritized combinatorial
testing process. Line 2 derives a reference output by executing the seed
program $P$ with the default configuration $\Gamma$. %
Line 3 retrieves an operator set
$Ops$ that records all operators in $P$. The function
\textsc{GetRelevantParameters} then prunes the full configuration space by
only selecting parameters relevant to the operators used in $P$,
particularly the adaptation strategies. Based on this refined scope, line 5
enumerates all pairs of configurations and orders them according to the Hamming
distance from previously selected configurations, then selects the top $N_{max}$
configurations to form a prioritized test suite $ParamList$, where $N_{max}$ is a
user-defined budget to limit the total number of tests. For each variant
configuration, \tool\ executes the program under $\Gamma'$ and reports any crashes or inconsistencies (lines
6--11).

\subsection{MR3: Backend and Scheme Mutation}
\label{subsec:mr3}

\parh{Design Goal.} This MR addresses Challenge III
(\S~\ref{subsec:motivation-challenge3}) by validating the framework's correctness
across different cryptographic and hardware components. The FHE ecosystem is
inherently heterogeneous, characterized by a rich diversity of cryptographic
schemes and backend implementations. This MR generates test cases by keeping the
program $P$ constant while mutating the backend execution environment $\beta$ or
the cryptographic scheme $\sigma$. The testing oracle asserts that the program's
output should remain invariant regardless of the underlying execution engine.
Specifically, we target three distinct layers of cross-stack inconsistencies by
systematically configuring the FHE framework to execute the same program $P$
under different settings, including hardware accelerators (such as CPU vs. GPU),
backend libraries (including OpenFHE vs. SEAL), and target FHE schemes (such as BFV
vs. BGV).

\parh{Cross-Stack Consistency Checking.} As described in
\S~\ref{subsec:motivation-challenge3}, defining a universal oracle for all these
schemes and backends is inherently challenging due to their heterogeneous
nature. To address this, \tool\ employs an adaptive tolerance-based oracle that
adjusts its strictness based on the settings being compared. When comparing
settings involving exact integer arithmetic only, such as
BGV vs. BFV, or CPU vs. GPU on integer logic, the oracle enforces strict
equality, ensuring $\mathcal{F}(P, \Gamma) = \mathcal{F}(P, \Gamma')$. However,
for settings involving approximate arithmetic, such as the CKKS scheme, inherent floating-point and approximation deviations
between backends are expected. In these cases, \tool\ relaxes the equivalence
condition to a tolerance-based check, verifying that the divergence falls within a
statistically acceptable error bound $\epsilon$. We set $\epsilon$ to 1\%
relative error based on analyzing benign floating-point deviations across 1,000
test cases with known correct outputs, where observed deviations consistently
remained below 0.5\%. As empirically validated in
\S~\ref{subsec:eval_component}, this approach effectively 
exposes significant semantic divergences indicative of framework bugs.

\noindent\textbf{Putting it all Together.} The three MRs are complementary and
collectively provide comprehensive coverage of the FHE framework stack. MR1
stresses data representation and noise management by mutating program structure,
uncovering bugs in data layout transformations and computational depth tracking.
MR2 explores the vast configuration space, exposing bugs in adaptation
strategies and optimization interactions. MR3 validates cross-stack consistency,
detecting semantic divergences across heterogeneous backends and schemes.
Together, these MRs systematically exercise the critical components identified
in \S~\ref{sec:motivation}, enabling \tool\ to detect both silent logic errors and
explicit crashes across the entire FHE framework pipeline. In real usage,
\tool\ iteratively generates seed programs and applies each MR to the seeds, enabling fine-grained defect detection
across diverse FHE framework implementations and configurations.

\section{Implementation and Experiment Setup}
\label{sec:implementation}

\parh{Implementation.} We implemented \tool\ primarily in Python, with approximately 3,300 lines of code
(LOC). All experiments were conducted on a server equipped with an AMD Ryzen
3970X 32-Core Processor, 256GB of RAM, and an NVIDIA GeForce RTX 3090 GPU. To
foster reproducibility and facilitate future research, we have open-sourced our
artifact at~\cite{anonymous_repo}. Key hyperparameters, such as the length of
seed programs, are calibrated based on a preliminary study of typical FHE
programming patterns and are documented in our artifact. 

\parh{Target FHE Frameworks.} We comprehensively reviewed current FHE frameworks
and selected three representative frameworks as our evaluation %
subjects: Google's HEIR, Zama's Concrete, and IBM's HELayers. These frameworks
are maintained by major industrial companies and have achieved significant
community traction. Furthermore, they
cover a diverse range of compilation strategies, cryptographic schemes, and
hardware backends. We tested the latest stable releases at the time of our
experiments: Concrete v2.11.0, HELayers v1.5.5.3, and HEIR v0.0.2.

HEIR is an MLIR-based FHE infrastructure offering a multi-level lowering
pipeline. In our experiments, we primarily target its Python frontend to evaluate
high-level compilation flows. Concrete specializes in the TFHE scheme and
supports multiple execution backends, including CPU and GPU. Notably, it provides
the most extensive support for high-level non-linear operators (such as
$\mathsf{sin}$, $\mathsf{cos}$) via LUTs. HELayers offers various interfaces
ranging from ONNX models to direct mathematical APIs. To ensure a fair and
consistent evaluation across all frameworks, we specifically target the
mathematical API layer, as it is the common interface supported by all three
frameworks. Nevertheless, the methodology of \tool\ is designed to be
input-agnostic and can be extended to support other high-level interfaces (such as
ONNX) in future work (further discussed in \S~\ref{sec:discussion}).

\section{Evaluation}
\label{sec:evaluation}

We evaluate \tool's effectiveness and analyze detected bugs using these
research questions:

\noindent\textbf{RQ1:} How effective and efficient
is \tool\ in detecting bugs across different frameworks?

\noindent\textbf{RQ2:} How do individual components in
\tool's testing pipeline contribute to bug detection?

\noindent\textbf{RQ3:} What root causes led to the defects of the FHE frameworks, and what lessons can be derived?

\noindent\textbf{RQ4:} What are the potential real-world
security implications of the detected bugs in FHE frameworks?

\begin{table}
\caption{Summary of bugs detected by \tool. ``Logic Errors'' refer to cases where the FHE framework completes execution but produces incorrect results. ``Crashes'' include obvious manifestations like assertion failures or memory errors.}
\label{tab:bug_summary}
\centering
\small
\resizebox{0.8\columnwidth}{!}{%
\begin{tabular}{c|c|cc}
\toprule
\multirow{2}{*}{\textbf{FHE Framework}} & \multirow{2}{*}{\textbf{Total Bugs}} & \multicolumn{2}{c}{\textbf{Bug Type}} \\
 & & \textbf{Logic Error} & \textbf{Crash} \\
\midrule
\textbf{Concrete} (Zama) & 15 & 12 & 3 \\
\textbf{HEIR} (Google) & 4 & 2 & 2 \\
\textbf{HELayers} (IBM) & 2 & 2 & 0 \\
\midrule
\textbf{Total} & \textbf{21} & \textbf{16} & \textbf{5} \\
\bottomrule
\end{tabular}%
}
\end{table}

\subsection{Overall Effectiveness and Efficiency}
\label{subsec:eval_overall}

\parh{Overall Effectiveness.}
\cref{tab:bug_summary} summarizes the bugs found by \tool\ during a continuous
testing campaign. 
We detected 21 previously unknown bugs across the three frameworks. Notably, 76.2\% (16/21) are logic errors, which can have critical implications and are
often more subtle and challenging to detect than crashes. Unlike crashes, logic
errors allow the framework to complete computation but yield incorrect results.
Triggering these errors often requires intricate operation sequences. For instance, certain logic bugs in
Concrete only manifest when an encrypted input is placed at a specific index
within an array and subsequently reused, or when arithmetic multiplication
interacts with specific non-linear operations (such as \texttt{maximum}) in a
precise order. Such errors are subtle as they do not lead to immediate crashes.
Due to the privacy-preserving nature of FHE, incorrect computations may go
undetected in privacy-sensitive applications and cause severe consequences such
as incorrect medical diagnoses or financial model divergence.

A higher concentration of bugs is observed in Concrete (15/21), despite the
similar testing effort across all frameworks. This distribution is expected and
can be attributed to the framework's richer feature set and diverse optimization
pipeline. Unlike other frameworks that support a limited set of arithmetic
operations, Concrete supports a rich set of NumPy-compatible operators, ranging
from standard arithmetic to complex bitwise logic and non-linear functions like
cosine and logarithm. Furthermore, it exposes a diverse set of configurations to
optimize for the underlying TFHE scheme. While these features significantly
enhance performance and usability, they also introduce a larger attack surface
for latent bugs.

In contrast, the current versions of HELayers' mathematical layer and HEIR offer
limited support for diverse non-linear functions and complex configurations.
Bugs in these frameworks were primarily triggered by our diverse FHE-aware high-level programs and cross-stack consistency checks,
identifying fundamental inconsistencies in how high-level arithmetic is lowered
to cryptographic backends like SEAL and OpenFHE. This distribution reflects the
current developmental stage of frameworks like HEIR, where frontend diversity
and arithmetic complexity are still evolving. The comprehensive results on
Concrete demonstrate \tool's capability to provide continuous integration
testing for all FHE frameworks as they mature and expand their operator support.

To ensure validity, all detected bugs were independently triaged by two experts
in FHE frameworks (also authors of this paper), who reached consensus that all
21 bugs are indeed true positives. 
We have responsibly reported all our findings to FHE framework developers, and several issues have been positively acknowledged with patches actively being developed~\cite{heir-bug1}.

\parh{Validity of Seeds.} The effectiveness of MT relies
heavily on the quality of the initial seed programs. If generated seeds are
syntactically invalid or fail to compile due to trivial errors, subsequent
mutations become futile. However, generating valid programs in the FHE context
presents unique challenges beyond conventional ``syntactic correctness.'' FHE
frameworks often impose strict validity constraints driven by performance
optimizations or underlying scheme limitations. For instance, Concrete rejects
intermediate values that exceed specific bit-widths, while HELayers aborts
computations if accumulated noise surpasses the decryption threshold.
Thus, a program can be syntactically correct yet cryptographically
invalid.

To balance generation efficiency with implementation complexity, we prioritized
testing throughput over guaranteeing validity via expensive static analysis of
cryptographic properties. Instead of employing heavyweight dataflow analysis to
predict intermediate values or noise budgets, we adopted a lightweight iterative
regeneration strategy. If a generated seed violates cryptographic constraints
during the dry-run, \tool\ discards it and attempts to regenerate a compliant
variant. With a maximum of five retry iterations, our approach achieves an
average validity rate of 98.93\%, including those that successfully trigger
framework vulnerabilities during the self-checking (\S~\ref{subsec:seed_generation}).
Among the small number of invalid seeds, approximately 23\% were discarded due
to execution timeouts, while the remainder failed due to intractable
cryptographic parameter unsatisfiabilities. We excluded these computationally
expensive cases for testing efficiency. The high validity rate demonstrates that
our strategy practically and effectively navigates the strict FHE constraints,
allowing the testing budget to focus on stressing deep logic rather than trivial
validity issues.

\begin{table}[t]
\caption{Efficiency statistics of \tool. ``Avg. Gen. Time'' represents the average time to generate a test case. ``Avg. Latency'' reports time overhead per test case, which is dominated by FHE compilation and execution. ``Throughput'' reflects the end-to-end testing speed.}
\label{tab:efficiency_metrics}
\centering
\small
\resizebox{\columnwidth}{!}{%
\begin{tabular}{l|c|c|c|c}
\toprule
\multirow{2}{*}{\textbf{Framework}} & \multirow{2}{*}{\textbf{Valid Rate}} & \textbf{Avg. Gen.} & \textbf{Avg. Latency} & \textbf{Throughput} \\
 & & \textbf{Time}  & \textbf{Compile \& Execute} & \textbf{(cases/core $\cdot$ h)} \\
\midrule
\textbf{Concrete} & 96.9\% & 1.87 ms & 189.46 s & 18.99 \\
\textbf{HEIR} & 100.0\% & 0.38 ms & 21.12 s  & 170.72 \\
\textbf{HELayers} & 99.9\% & 1.24 ms &  5.20 s & 690.72 \\
\midrule
\textbf{Average} & 98.93\% & 1.16 ms & 71.93 s & 293.48 \\
\bottomrule
\end{tabular}%
}
\end{table}

\parh{Throughput.}
Testing FHE frameworks faces the inherent challenge of expensive FHE
computations, which are often orders of magnitude slower than plaintext
execution. Despite this challenge, \tool\ maintains practical testing
throughput, generating and processing an average of 293.48 test cases per core
hour. As detailed in \cref{tab:efficiency_metrics}, seed generation time is
negligible, with primary latency stemming from compilation and execution phases.
The variance in
throughput across frameworks primarily reflects differences in the underlying
cryptographic schemes and operator complexity---Concrete's TFHE scheme with
extensive operator support incurs higher costs than the more streamlined
implementations in HEIR and HELayers. For our large-scale
evaluation involving 3,000 seed programs, the entire testing campaign consumed an
average computational cost of approximately 303.7 core-hours per framework.
While the intrinsic characteristics of FHE operations limit absolute testing
speed compared to traditional compiler fuzzing, the high bug detection count
shown in \cref{tab:bug_summary} validates the efficiency of our design. Our
crafted seed generation and targeted MR designs ensure that each executed test
case is highly effective, enabling the discovery of hidden logic errors
within a reasonable computational budget.

\parh{Summary.} In summary, our evaluation demonstrates that \tool\ achieves
high effectiveness (21 previously unknown bugs across three major frameworks) with
practical efficiency (293 test cases per core hour on average), while
maintaining a high seed validity rate (98.93\%). The predominance of logic
errors ($\frac{16}{21}=76.2\%$) underscores the importance of our MT approach, as these silent bugs are particularly challenging to detect
and can have severe real-world consequences in privacy-sensitive applications.

\subsection{Effectiveness of Individual Components}
\label{subsec:eval_component}

\begin{table}[!htbp]
\caption{Distribution of unique bugs detected by each component of \tool. ``FHE-aware Seed
Gen.'' denotes bugs triggered by the initial seed alone, while MRs indicate bugs
revealed only after specific mutations.}
\label{tab:component_contribution}
\centering
\small
\resizebox{0.85\columnwidth}{!}{%
\begin{tabular}{c|c|c|c|c}
\toprule
\multirow{2}{*}{\textbf{Component}} & \textbf{FHE-aware}  & \textbf{MR1} & \textbf{MR2}  & \textbf{MR3} \\
& \textbf{Seed Gen.} & \textbf{Program} & \textbf{Config.} & \textbf{Backend} \\
\midrule
Bug Count 		    & 12 & 5 & 1  & 3 \\
\bottomrule
\end{tabular}%
}
\end{table}

\begin{figure}[t]
  \centering
  \lstset{
    language=Python,
    basicstyle=\ttfamily\scriptsize,
    frame=single,
    numbers=none,
    xleftmargin=0pt,
    framesep=2pt,
    aboveskip=1pt,
    belowskip=1pt,
    breaklines=true,
    columns=fullflexible
  }

  \begin{subfigure}[t]{0.48\columnwidth}
\begin{lstlisting}
x = hint(x, bit_width=8)
y = hint(y, bit_width=16)
z = np.minimum(x, y)
return z + y
\end{lstlisting}
    \caption{Logic bug triggered by specific bit-width combination.}
    \label{fig:case_mr1}
  \end{subfigure}
  \hfill
  \begin{subfigure}[t]{0.48\columnwidth}
\begin{lstlisting}
t = y - 12
t = (9 == t) * 1
t = (t - 1) * 8
return t * -10
\end{lstlisting}
    \caption{Discrepancy bug between CPU and GPU backends.}
    \label{fig:case_mr3}
  \end{subfigure}

  \caption{Representative bug cases found by different MRs.}
  \label{fig:ablation_cases}
\end{figure}

To verify the necessity of our multi-layered testing strategy, we analyze the
distinct contributions of the FHE-aware seed generator and each MR. As detailed in
\cref{tab:component_contribution}, which presents the distribution of unique
bugs detected by each component, the seed generator accounts for about half of the
findings. This confirms that our generation strategy, prioritizing high
multiplicative depth and diverse non-linear functions, successfully covers edge
cases often overlooked by FHE framework developers, laying a robust foundation
for subsequent metamorphic mutations. However, this does not diminish the value of our MRs; 
on the contrary, the following analysis reveals that the MRs target
orthogonal classes of vulnerabilities that are unreachable by seed generation
alone.

MR1 is indispensable for detecting logic bugs, %
particularly those related to data types and complex arithmetic rewriting. While
the seed generator creates valid syntax, it often defaults to standard types and
structures. \tool's MR1 proactively mutates data types and structure layouts via
semantically equivalent transformations. This capability was crucial for detecting
the array-related logic bug shown in \cref{fig:mot_concrete_array} and the
critical memory safety violation discussed in \S~\ref{subsec:security_implication}. These
vulnerabilities were exposed solely by MR1's layout mutations, as they require
specific structural perturbations that are difficult to achieve by the seed
generator alone.

To further illustrate MR1's effectiveness, we present a case study regarding type
mutation in \cref{fig:case_mr1}. This logic bug produces output drastically
different from the expected result
when input $y$ is mutated to a 16-bit width. 
For instance, given inputs $x=1$ and $y=2$, the program may output random 5-digit integers (such as $19796$) instead of the expected 3. 
Our analysis of the
intermediate representation of the compiled circuit showed no anomalies,
indicating the flaw lies deep within the underlying backend's handling
of mixed-precision operations. Without MR1's targeted type mutation, such
backend defects are difficult to expose systematically.

MR3 proves critical for verifying consistency across different computing
backends and encryption schemes. \cref{fig:case_mr3} illustrates a simplified
representative case demonstrating the synergy between our seed generator and
MR3. The generated program, involving comparison operations and scalar
arithmetic, executes correctly on the CPU backend. However, when MR3 switches
to the GPU backend, the program yields unstable and incorrect results.
Through systematic minimization, we found that the bug disappears when either the multiplicative depth is reduced or the comparison operation is removed. This
suggests a flaw in how the GPU backend handles the lowering of comparison
operators under deep multiplicative chains. This case highlights that the
complex logic generated by our seed generator requires MR3's cross-backend
validation to expose backend-specific implementation gaps that would otherwise
remain latent.

While MR2 detected relatively fewer direct bugs, the nature of these discoveries
is significant. The bug shown in \cref{fig:mot_concrete_config} demonstrates a
critical availability threat: an assertion failure triggered exclusively by a
specific, valid configuration choice. Such configuration-dependent failures are
particularly insidious as they represent hidden vulnerabilities that only
manifest under specific optimization settings, making them harder to detect than
standard crashes. Beyond confirmed bugs, MR2 also revealed numerous instances of
``noise overflow runtime errors'' during configuration transitions. Although we
conservatively excluded these from our bug count as they may represent
legitimate parameter unsatisfiability, they highlight a significant usability
gap---ensuring consistent behavior across the vast configuration space
remains a challenge for current FHE frameworks. The relatively lower bug count
for MR2 also reflects the current stage of FHE frameworks, which %
are still evolving to support diverse arithmetic operators and adaptation
configurations. As these frameworks mature and expand their configuration
spaces, we anticipate that MR2's systematic exploration of configuration
interactions will become increasingly valuable.

To further demonstrate the necessity of our FHE-aware design, we compare \tool\ with Domato~\cite{domato}, a representative fuzzer from Google. While this grammar-based fuzzer has proven to find thousands of critical bugs in complex software, it lacks FHE-specific generation and mutation strategies. When evaluated on Concrete with the same testing budget (see Appendix~\ref{app:baseline} for details), Domato finds only 3 crash-triggering inputs and no logic errors. In contrast, \tool’s FHE-aware seed generation finds 7 logic-error inputs and 7 crash inputs, while our MRs further uncover 6 additional logic-error inputs and 18 crash inputs. These results show that FHE-aware seed generation reaches FHE-specific failure paths more effectively than generic grammar generation, and that our MRs reveal complementary failures beyond seed generation.

\begin{table}[t]
\centering
\setlength{\tabcolsep}{2.5pt}
\caption{Root-cause families of the 21 defects discovered by HERTA.
L/C denotes logic errors and crashes. R1--R5 are
FHE-specific; R6 captures standard robustness defects.}
\label{tab:root-causes}
\begin{tabular}{c|c|p{0.7\columnwidth}}
\toprule
\textbf{Family} & \textbf{\#(L/C)} & \textbf{Root Cause Family Description} \\
\midrule
R1 &
7 (6/1) &
Incorrect representations for scalars. \\

R2 &
3 (2/1) &
Incorrect lowering for operators.
\\

R3 &
3 (3/0) &
Wrong value packing or materialization. \\

R4 &
3 (2/1) &
Inconsistent or overly conservative selection of cryptographic parameters. \\

R5 &
3 (3/0) &
Faulty backend or scheme lowering. \\

R6 &
2 (0/2) &
Programs cause aborts or memory corruptions. \\
\bottomrule
\end{tabular}
\end{table}

\subsection{Analysis of Root Causes}
\label{sec:root-causes}

\modify{
We analyze root causes of our uncovered bugs using
minimized triggers, observed failures, and available developer feedback.
The 21 bugs are categorized into 6 root-cause families, as summarized in \cref{tab:root-causes}. Out of these, R1--R5 (19 bugs in total) are specific to FHE framework design and implementation, while R6 captures robustness issues that also arise in general software systems.

\noindent\textbf{R1: Incorrect Scalar Representation.}
These bugs stem from mismatches between source-level scalar semantics and FHE number representations. For instance, HEIR mishandles scalar signedness during lowering: a negative constant, when used as a multiplicand for an encrypted variable, is interpreted as unsigned during bit-width extension; another Concrete bug is triggered simply by widening one operand's declared bit-width, with runtime values unchanged.

\noindent\textbf{R2: Incorrect Lowering for Operators.}
This type of bug miscompiles combinations of operators under specific contexts. 
Concrete, for instance, triggers incorrect optimizations for programs involving multiplication and \texttt{maximum} operators in a particular order. Another program with a non-default configuration, \texttt{THREE\_TLU\_CASTED}, crashes due to incorrect lowering for the min/max operators. These cases do not arise from a single operator, but rather from the interaction between operator adaptation, computation, and configuration choices, and thus are hard to detect without FHE-specific testing.

\noindent\textbf{R3: Errors in Data Packing and Materialization.}
To optimize performance, FHE frameworks may pack multiple numbers into encrypted slots. Bugs in this category arise when frameworks fail to correctly bundle this data or format the final decrypted outputs. For instance, the Concrete framework corrupts computations when embedding an encrypted number into an array and later pulling it back out. In another case, a variable yielded different results depending on whether it was stored alone or alongside others. These issues highlight flaws in how frameworks translate between user variables and internal packed formats.

\noindent\textbf{R4: Inaccurate Noise Management and Parameter Synthesis.}
A unique challenge in FHE is managing the ciphertext noise that grows with each operation (multiplicative depth). Frameworks must automatically synthesize cryptographic parameters large enough to prevent this noise from corrupting the final result. Bugs in this category happen when frameworks misjudge this ``noise budget.'' For example, Concrete underestimated the parameters needed for deep arithmetic and comparisons, leading to silently corrupted decrypted values. Conversely, HEIR's noise validation was overly conservative, falsely rejecting a trivially simple and valid BGV program.

\noindent\textbf{R5: Cross-Backend and Scheme Inconsistencies.}
FHE frameworks offer developers the flexibility to switch between different
cryptographic backends or schemes for performance or compatibility reasons.
However, we found bugs where changing this execution stack silently breaks
consistency. For instance, in Concrete, a program yielded the correct result on
a CPU but produced a completely different answer when run on a GPU. Similarly,
in HELayers, identical computations diverged significantly between the OpenFHE
and SEAL libraries, as their output differences exceeded acceptable
approximation noise. These errors emphasize that semantic equivalence across the
FHE ecosystem remains fragile.

\noindent\textbf{R6: Runtime Robustness and Memory Safety.}
Unlike the previous categories that cause silent data corruption, R6 manifests as explicit crashes. We found that certain program constructs bypass safe memory management within the framework. For instance, Concrete suffered fatal heap metadata corruption when an encrypted variable was placed at the end of an array. While these bugs do not corrupt the mathematical logic, they represent critical security vulnerabilities that can crash backend servers in FHE-as-a-Service environments.

\noindent\textbf{Design Implications.} To guide future FHE framework development, we highlight key design implications by reflecting on bugs we found. By rethinking how frameworks manage layout, noise, and backend lowering, FHE framework developers can proactively eliminate major sources of silent data corruption.

\emph{Lesson 1: Explicit Representation Validation.} FHE frameworks
often translate program representations across multiple layers, from
source-level data types and IRs to ciphertext encodings and packed backend
layouts. This complexity can allow subtle bugs in data representation and
packing to silently corrupt computations, as demonstrated by R1 and R3.
Frameworks should encode these representation properties as explicit IR
contracts and check them after passes that change layouts or types so that developers can detect miscompilations caused by representation errors early.

\emph{Lesson 2: Transparent Noise Management.} Current FHE frameworks treat
parameter synthesis and noise budget estimation as black boxes; while this abstraction hides complexity, it also hides critical information from users,
potentially causing silent data corruption or false rejections. Instead of
relying solely on brittle ahead-of-time estimations, frameworks can provide
built-in diagnostics. By implementing an intermediate value inspection
mode that dynamically tracks noise growth during debugging, frameworks can embed
runtime assertions that warn users when computations approach theoretical
thresholds to prevent the silent decryption of garbage data.

\emph{Lesson 3: Strict Semantic Contracts for Backend Retargeting.} Frameworks
often rely on implicit backend-specific adaptation and lowering strategies for
different hardware platforms or FHE libraries. This can cause semantic drift
across backends and confuse users. Frameworks should embed explicit semantic
contracts into their IRs, ensuring that retargeting to a GPU or a different
library remains semantics-preserving, or that unsupported operations are
rejected with clear diagnostics.
}

\subsection{Security Implications of Bugs}
\label{subsec:security_implication}

In this section, we conduct a hazard analysis to demonstrate how the latent bugs
detected by \tool\ can be weaponized by adversaries to facilitate stealthy
attacks on real-world privacy-preserving applications. As described in
\S~\ref{subsec:motivation_significance}, FHE is increasingly adopted in
high-stakes, data-sensitive environments, such as genomic privacy analysis and
decentralized financial auditing. Logic bugs in FHE frameworks can silently lead to incorrect medical diagnoses or financial model divergence, misleading clinicians to mistreat patients or causing financial risks to go undetected. We illustrate three representative attack scenarios that leverage our detected bugs in FHE frameworks:

\noindent\textbf{Integrity Breach: Subverting Source Code Review via Framework Bugs.}
In privacy-sensitive applications, developers heavily rely on source-level code
reviews to ensure computational integrity. However, FHE frameworks can be
exploited to create a severe gap between what the code appears to do and what it
actually executes. A malicious contributor to an FHE application can craft code
that is mathematically sound to code reviewers but triggers a framework-level %
lowering bug.

For example, we discovered a bug in HEIR (\cref{fig:heir_attack}) where a signed -1 scalar is improperly lowered as a large unsigned integer (e.g., 255 for 8-bit representations) when multiplied with an encrypted variable. To an auditor, \(-1 \times liability\) is mathematically equivalent to subtracting \(liability\). However, due to the bug, the compiled encrypted computation adds an amplified liability term instead of subtracting it. This allows an adversary to manipulate the net-worth score without needing malformed ciphertexts or decryption keys, effectively breaking computational integrity at the framework layer while passing code review. In total, we found 4 similar attack patterns, where signed scalar lowering, mixed bit-width handling, and max/min operator lowering bugs can be exploited to bypass source-level audits and cause integrity breaches in high-stakes FHE applications like encrypted scoring and risk assessment.

\begin{figure}[!htbp]
\centering
\begin{lstlisting}[language=python, basicstyle=\ttfamily\scriptsize, xleftmargin=1em, frame=single]
def secure_net_worth(asset_bank_a: Secret[I32], 
                     asset_stock_b: Secret[I32], 
                     liability_loan_c: Secret[I32]):
    total_assets = asset_bank_a + asset_stock_b
    deduction = -1 * liability_loan_c   # Miscompiled line
    net_worth = total_assets + deduction
    return net_worth
\end{lstlisting} 
\caption{An example of framework-assisted backdoor attack where the compiled circuit executes an amplified addition instead of the intended subtraction.} 
\label{fig:heir_attack} 
\end{figure}

\noindent\textbf{Audit Evasion: Exploiting Backend Inconsistencies.}
Modern FHE frameworks often expose backend retargeting as a deployment-time
optimization. A developer may validate an FHE program on a reference backend
during development or audit, and later switch to another backend library or
hardware target for performance, compatibility, or cost. Attackers can exploit framework inconsistencies across these backends to bypass security audits.

We identified a program in Concrete (\cref{fig:case_mr3}) that executes correctly on a CPU but produces corrupted results when retargeted to a GPU, due to a flaw in how the GPU backend lowers comparison operators under deep arithmetic chains. An adversary can introduce this specific program structure into the source code, which will pass code review and reference-backend testing. However, once deployed to the production GPU for performance, the program silently evaluates a different result, potentially leading to a manipulated financial score. We found 3 similar cases, such as discrepancies between SEAL and OpenFHE backends in HELayers. These bugs effectively render reference audits obsolete and turn performance-oriented backend retargeting into a critical security risk.

\noindent\textbf{Availability Breach: Valid Program Denial-of-Service.}
The computational intensity of FHE has driven the emergence of
``FHE-as-a-Service''~\cite{GraphAIHEaaS, yadavalli2025homomorphic}, allowing
users to compile and deploy their FHE-enhanced applications on managed cloud
platforms~\cite{MSHECloud, AWSHECloud}. In these shared environments, framework-level memory safety bugs expose the system to severe Denial-of-Service (DoS) attacks. 

We found 2 bugs where an adversary can submit syntactically valid FHE programs that easily pass frontend validation but trigger fatal compilation or execution errors. For instance, \tool\ uncovered a bug in Concrete where a valid array initialization pattern, involving an encrypted input as the final element, corrupts heap metadata during execution, leading to a \texttt{malloc: corrupted top size} error and crashing the process. In a cloud environment, an attacker can repeatedly submit such programs to crash cloud worker nodes and degrade service availability for legitimate users.

\noindent\textbf{Scenario Reflection.}
While these attack scenarios demonstrate serious security implications, we acknowledge several limitations in their practical exploitability. First, an integrity breach requires attackers to identify specific vulnerabilities and craft triggering code. Although reviewers might spot suspicious code patterns, like multiplying by negative scalars instead of subtracting, such code will become harder to distinguish from legitimate optimizations as FHE frameworks grow in complexity. Second, audit evasion relies on a mismatch between audited and production backends. While exhaustive validation of every backend and configuration could mitigate this, it is often prohibitively expensive. Since retargeting to GPUs or alternative libraries is routinely performed for throughput or cost, we believe this scenario represents a realistic risk. Third, regarding DoS attacks, cloud platforms may use rate limiting or input validation to mitigate repeated crashes. Nevertheless, the existence of these vulnerabilities, especially silent logic errors, represents a fundamental threat to the trustworthiness of privacy-preserving applications.

\section{Discussion}
\label{sec:discussion}

\parh{Alternative Testing Methods.} Besides MT, differential
testing (DT) is commonly used in compiler testing
efforts~\cite{polito2022interpreter, li2023finding, wang2024rustlantis}. In DT,
different compilers should produce identical outputs for the same input;
discrepancies indicate bugs. However, applying DT
to FHE frameworks presents unique challenges. First, the sophisticated FHE compilation and execution pipeline makes DT against
different frameworks insufficient for isolating root causes, especially in
complex cryptographic contexts. Second, the rapid evolution of
FHE frameworks introduces specialized internal APIs and primitives that lack
direct semantic equivalences in other frameworks, complicating
cross-domain comparisons. Third, and most crucially, FHE frameworks often employ
different approximation strategies, where valid
decrypted results inherently diverge from exact values. Distinguishing
between genuine bug-induced errors and acceptable approximation noise is
algorithmically difficult in a DT setting. In contrast, the MT
approach adopted by \tool\ enables fine-grained, self-contained, and efficient
testing that is better suited to the current FHE landscape.

\parh{Testing vs. Verification.} Although formal verification techniques provide
stronger correctness guarantees by rigorously proving specific properties, the
immense cryptographic complexity, the size of FHE framework codebases (often
containing hundreds of thousands of lines of code~\cite{Concrete, ali2025heir}),
and the rapidly evolving nature of modern FHE frameworks make constructing
comprehensive formal models prohibitively expensive. In line with other work on
software reliability in complex systems~\cite{aumasson2017automated,
somorovsky2016systematic}, we adopt automated testing as a pragmatic approach.
It offers a scalable solution for detecting bugs in real-world FHE framework
implementations, facilitating iterative improvements and rapid developer feedback. 
The reported issues can serve as concrete inputs for developers
to debug and patch their frameworks. We leave formal verification of FHE
frameworks as a complementary future direction.

\parh{Limitations of Language Feature Support.} We acknowledge that \tool\
primarily focuses on the fundamental arithmetic layers of FHE frameworks and
does not cover every high-level, domain-specific API offered by certain
frameworks, such as the ONNX input interface in HELayers. However, the
correctness of these high-level abstractions fundamentally relies on the
reliability of the underlying cryptographic primitives. By targeting these core
primitives, our MRs effectively test the foundational logic
where critical failures often originate. Furthermore, given that FHE frameworks
implement distinct and rapidly evolving interfaces, maintaining a comprehensive
seed generator that covers all possible APIs presents a significant maintenance
challenge. Nevertheless, our evaluation demonstrates that \tool's current
strategy effectively uncovers a substantial number of bugs across multiple
layers of the FHE stack. We therefore leave the development of a dynamic seed
generator capable of automatically adapting to evolving FHE APIs as a promising
direction for future work.

\section{Related Work}
\label{sec:related}

\parh{Security Assurance for Cryptographic Libraries.} Securing cryptographic
software demands a multi-faceted approach.
DY fuzzer~\cite{10646658} and TLS-Attacker~\cite{somorovsky2016systematic}
support automated analysis and testing of cryptographic protocol implementations. 
TaintCrypt~\cite{rahaman2021theory} uses taint analysis to find logical
bugs in C/C++ code, while CryptoGuard~\cite{rahaman2019cryptoguard} detects
cryptographic API misuse in Java projects. CryptoLine~\cite{fu2019signed}
formally verifies foundational cryptography in assembly. Unlike these works that
focus on traditional cryptographic libraries, \tool\ addresses the unique
challenges of testing FHE frameworks, which involve a complex 
pipeline from high-level code to low-level cryptographic operations. Testing
such frameworks presents distinct challenges due to the cryptographic properties
of FHE (such as noise accumulation and approximate arithmetic) and the heterogeneous
nature of frameworks' compilation and execution stacks.

\parh{Vulnerability Detection in Compilers.} Compilers translate high-level
code into low-level machine instructions. Their complexity makes them prone to bugs that compromise software security~\cite{xu2023silenta}. While extensive
work targets JavaScript JIT compilers~\cite{wang2024optfuzz, xu2024fuzzing,
wachter2025dumpling, gross2023fuzzilli}, C/C++ compilers~\cite{yang2011finding,
le2014compiler}, and deep learning compilers~\cite{liu2023nnsmith,
zhang2025deep}, \tool\ addresses the unique challenges of FHE frameworks. It
introduces novel domain-specific MRs to probe FHE-intrinsic
properties such as noise management, parameter selection, and approximate
arithmetic. Furthermore, unlike traditional DT that relies on
cross-compiler comparisons, \tool\ navigates the heterogeneous configuration
space within a single framework to validate semantic consistency across
different backends, cryptographic schemes, and adaptation strategies.

\section{Conclusion}
\label{sec:conclusion}

We introduce \tool, a systematic testing framework for FHE framework correctness that leverages FHE-specific MRs. \tool\ uncovered 21 previously unknown bugs in three mainstream frameworks, including critical logic and memory safety errors. We release \tool\ to support building more reliable privacy-preserving systems.

\bibliographystyle{IEEEtran}
\bibliography{bib/main}

\appendix

\section{Appendix}
\label{sec:appendix}

\subsection{Comparison with a Generic Fuzzing Baseline}
\label{app:baseline}

To further isolate the benefit of \tool's FHE-aware design, we compare \tool\ with a grammar-based fuzzer, Domato, on Concrete. Both tools use the same operator set, input domains, maximum program length, and 1K valid seed budget. Unlike \tool, Domato does not use FHE-aware heuristics such as multiplicative-depth prioritization.

\begin{table}[!htbp]
\centering
\caption{Comparison with Domato under 1K valid seed budget on Concrete. \tool\ (Seed Only) uses only the FHE-aware seed generator, while \tool\ (Seed+MRs) includes the generated seeds and their MR-generated variants. Values in parentheses indicate the additional failure-triggering inputs exposed by MRs beyond those found by seed only. \#Crash Inputs and \#Logic-Error Inputs denote the numbers of test cases that trigger crashes and logic errors, respectively.}
\label{tab:domato_baseline}
\small
\begin{tabular}{lrr}
\toprule
\textbf{Method} & \textbf{\#Crash Inputs} & \textbf{\#Logic-Error Inputs} \\
\midrule
Domato & 3 & 0 \\
\tool\ (Seed Only) & 7 & 7 \\
\tool\ (Seed+MRs) & 25 (+18) & 13 (+6) \\
\bottomrule
\end{tabular}
\end{table}

\cref{tab:domato_baseline} shows that generic grammar generation exercises basic frontend paths, but is less effective at reaching FHE-specific failure modes under the same valid-seed budget. In particular, Domato triggers only three crashes and no logic errors, while \tool's seed generation exposes both crashes and silent logic errors. \tool's MRs further expose additional failure-triggering inputs, showing that \tool's FHE-aware seed generation and MRs provide complementary benefits over generic seed-only fuzzing.

\end{document}